\documentclass{nature}
\usepackage{amsmath, amssymb, dsfont}
\usepackage{graphicx}

\title{Quantum Anomalous Hall Effect in Magnetic Topological Insulator GdBiTe$_3$}

\author{Hai-Jun Zhang, Xiao Zhang  \& Shou-Cheng Zhang}
\begin{document}

\maketitle

\begin{affiliations}
\item Department of Physics, McCullough Building, Stanford University, Stanford, CA 94305-4045
\end{affiliations}

\begin{abstract}
{\bf The quantum anomalous Hall (QAH) state is a two-dimensional bulk insulator with a non-zero Chern number in absence of external magnetic fields. Protected gapless chiral edge states enable  dissipationless current transport in electronic devices. Doping topological insulators with random magnetic impurities could realize the QAH state, but magnetic order is difficult to establish experimentally in the bulk insulating limit. Here we predict that the single quintuple layer of GdBiTe$_3$ film could be a stoichiometric QAH insulator based on \emph{ab-initio} calculations, which explicitly demonstrate ferromagnetic order and chiral edge states inside the bulk gap. We further investigate the topological quantum phase transition by tuning the lattice constant and interactions.  A simple low-energy effective model is presented to capture the salient physical feature of this topological material.}
\end{abstract}

Recently time-reversal-invariant topological insulators(TIs) have attracted broad attention in fields of the condensed matter physics, material science and electrical engineering\cite{qi2010a,Hasan2010,Moore2010,Qi2010b}. Quantum spin Hall(QSH) insulators\cite{bernevig2006d,koenig2007} are two dimensions(2D) TIs with a bulk energy gap and gapless helical edge states which are protected by the $Z_2$ topological invariant\cite{kane2005b}. Both 2D and 3D TIs are natural platforms to realize the QAH when the time-reversal symmetry is spontaneously broken by the ferromagnetic order\cite{qi2006,liu2008,Yu2010,qi2008a}. The QAH
state is the quantum version of the intrinsic anomalous Hall effect \cite{Nagaosa2010}, and is characterized by a bulk 2D Chern number and gapless chiral edge states without the external magnetic field\cite{haldane1988,qi2006,Onoda2003}, and is sometimes also referred to as the 2D Chern insulator. The QSH state has two counter-propagating edge states; doping the 2D TIs with magnetic impurities could break the time reversal symmetry, annihilate one edge state and leave the other intact, thus realizing the QAH state with chiral edge state. Alternatively, doping the surface of 3D TIs with magnetic impurities breaks time reversal symmetry, and opens up a gap for the Dirac surface states\cite{qi2008a,liu2009,chen2010}. Chiral edge states exist on the magnetic domain walls on the surface\cite{qi2008a}. In the QAH state, electrons move like cars on a highway, where oppositely moving traffics are spatially separated into opposite lanes\cite{qi2010a}. Realizing such a dissipationless transport mechanism without extreme condition could greatly improve the performance of electronic devices.

Normally, ferromagnetic order among magnetic impurities in randomly doped semiconductors are mediated by the free carriers \cite{Dietl2000}, it is difficult to establish magnetic order in the bulk insulating limit. For this reason, we search for stoichiometric ferromagnetic insulators with a nonzero Chern number for the 2D band structure. Rare earth elements have partially filled \textbf{f} electrons, in particular, Gd$^{3+}$ has exact half filled \textbf{f} electrons. It is possible to realize a stoichiometric QAH insulator by replacing Bi by Gd in the Bi$_2$Te$_3$ family of TIs\cite{zhang2009,Xia2009,Chen2009,yan2010}. GdBiTe$_3$ has a stable structure phase, synthesized thirty years ago\cite{spring2011}. In this work, we predict that stoichiometric QAH insulators could be realized in the single quintuple layer(SQL) of GdBiTe$_3$, based on {\it ab-initio} calculations.

\textbf{Crystal structure and band structure}

GdBiTe$_3$ has a rhombohedral crystal structure with the space group $R\overline{3}m$. Its crystal consists of close-packed layers stacked along [111] direction with the A-B-C$\cdots$ order. The quintuple layer (QL) (Te-X-Te-X-Te) structure is the basic crystal unit, where X presents Bi and Gd with a certain pattern, similar to the case of LaBiTe$_3$\cite{yan2010}. It has strong coupling within one QL and weak van der Waals coupling between neighbor QLs, so this material can be grown as two-dimensional (2D) thin film along [111] direction. In addition, though its lattice type and lattice constant have been measured experimentally\cite{spring2011}, positions of Bi and Gd have not been clearly resolved yet. Since Bi and Gd have different electronegativity, it is possible to grow the non-mixing structure as Te-Gd-Te-Bi-Te with Molecular beam epitaxy (MBE) technique. This crystal structure of GdBiTe3 is shown in Fig.~1a. It holds the three-fold rotation symmetry $C_3$ with $z$ axis as the trigonal axis and the reflection symmetry with $x$ axis as its normal axis. Compared with Bi$_2$Te$_3$\cite{zhang2009}, the inversion symmetry is broken for this structure. Since SQL GdBiTe$_3$ film is the simplest system to study the QAH effect, in this work we focus on its SQL structure with Te1-Bi-Te2-Gd-Te1$'$, where Te1 and Te1$'$ are marked for the Te closest to Bi and Gd layer separately, and Te2 denotes the Te in the center of this SQL.

All {\it ab-initio} calculations are carried out in the framework of density functional theory (DFT) with Perdew-Burke-Ernzerbof-type generalized-gradient approximation\cite{perdew1996}. Both BSTATE package\cite{fang2002} with plane-wave pseudo-potential method and the {\it Vienna ab initio simulation package} (VASP) with the projected augmented wave method are employed. The \textbf{k}-point grid is taken as $12\times12\times1$, and the kinetic energy cutoff is fixed to 340eV in all self-consistent calculations. A free standing slab model is employed with SQL. Its lattice constant ($a=4.16{\AA}$) is taken from experiments\cite{spring2011}, and the inner atomic positions are obtained through the ionic relaxation with the force cutoff 0.001eV/{\AA}. The spin-orbit coupling (SOC) is taken into account, because of its importance to realize the QAH effect. GGA+U method\cite{Anisimov1993} is also employed to study the strong correlation effect in GdBiTe$_3$ because of the existence of narrow occupied \textbf{f} bands of Gd.  We find that the ferromagnetic phase is more stable than the non-magnetic and collinear anti-ferromagnetic phases for SQL GdBiTe$_3$. In addition, the ferromagnetic phase with magnetic moment along [111] direction has lower energy than that with the magnetic moment along [010] and [110] directions. All GdBiTe$_3$ calculations are carried out here with the ferromagnetic moments along [111] direction.

The ferromagnetic phase of SGL GdBiTe$_3$ is an insulator state with exact magnetic moment $S=7/2$.  Because of the lattice similarity between GdBiTe$_3$ and GdN(or EuO) with [111] direction, the known magnetic exchange mechanisms in GdN or EuO could provide possible explanation for the ferromagnetic state in SQL GdBiTe$_3$.  The first mechanism is a third-order perturbation process\cite{mauger1986}. A virtual excitation, which takes a 4\textbf{f} to a 5\textbf{d} state, leads to a \textbf{f-f} interaction through the \textbf{d-f} exchange due to the wave-function overlap between neighboring rare-earth atoms. Recently, Mitra and Lambrecht\cite{mitra2010}, based on {\it ab-initio} calculations, presented another magnetic mechanism for the ferromagnetic ground state in GdN. The anti-ferromagnetic ordering, between N \textbf{p} and Gd \textbf{d} small magnetic moments, stabilizes the ferromagnetic structure between nearest neighbor Gd atoms due to the \textbf{d-f} exchange interaction. In addition, our calculations also indicate the similar anti-ferromagnetic ordering between Te \textbf{p} and Gd \textbf{d} small magnetic moments.

Bi$_2$Te$_3$, LaBiTe$_3$ and GdBiTe$_3$ have very similar SQL structure. First of all, Bi$_2$Te$_3$ have both the inversion symmetry and the time-reversal symmetry. However, inversion symmetry is broken in LaBiTe$_3$, and both inversion symmetry and time-reversal symmetry are broken in GdBiTe$_3$. The band evolution from Bi$_2$Te$_3$ to LaBiTe$_3$, and finally to GdBiTe$_3$ is shown in Fig.~1c-e. The bands of Bi$_2$Te$_3$ have double degeneracy  because of both inversion and time-reversal symmetries. Its energy gap is calculated to be about $0.19eV$, consistent with the experiments\cite{li2010}. The bottom of the conduction bands at $\Gamma$ point with a ``V'' shape originates from the $p_{x,y}$ orbitals of Bi, Te1 and Te1$'$. The top of the valence band at the $\Gamma$ point, which lies below valence bands at other momenta, originates from the $p_{x,y}$ orbitals of Te2. For LaBiTe$_3$, the double degeneracy of the bands is lifted except at the time-reversal points, due to the lack of the inversion symmetry. Though the bottom of conduction bands still shows the ``V'' shape, the top of the valence bands moves to $\Gamma$, which mainly originates from the $p_{x,y}$ orbitals of Te2 and Te1$'$. The schematic evolution of the top of valence bands and the bottom of conduction bands from LaBiTe$_3$ to GdBiTe$_3$ is shown in Fig. 2a-b. The Kramers double degeneracy at the $\Gamma$ point in GdBiTe$_3$ is broken due to the lack of time-reversal symmetry. The band inversion occurs between the $|p_{x}+\mathrm{i} p_{y},\mathrm{Bi+Te1},\downarrow\rangle$ and  $|p_{x}+\mathrm{i} p_{y},\mathrm{Te1'+Te2},\uparrow\rangle$ due to the strong SOC and the large magnetic moment. The existence of this band inversion is the key point to realize the QAH effect in this SQL GdBiTe$_3$ system.

Fig.~3a shows that the energy gap is quite sensitive with the lattice constant. Smaller lattice constant $a$ increases the band inversion. Conversely, increasing the lattice constant $a$ would destroy the band inversion. The QAH effect exists for $a<1.008a_0$, where $a_0$ is the experimentally determined bulk lattice constant. Because the SQL GdBiTe$_3$ is very thin, its lattice constant could be controlled by the substrate, giving us a broad tunability range. In addition, due to the narrow \textbf{f} bands, the strong correlation needs to be checked by GGA+U calculations. First of all, the band structure calculations show that the occupied \textbf{f} bands are located at $-6eV$ below Fermi level (FL), and the unoccupied \textbf{f} bands are located at $2.5eV$ above FL. Since both occupied and unoccupied \textbf{f} bands are quite far from FL, correlation effects should not influence bands very close to the Fermi level. Our calculations indicate that this conclusion is true. The dependence of the energy gap on U (from $0eV$ to $8eV$) is shown in Fig.~3b. The correlation U lifts up the top spin-down valence band, and pulls down the bottom spin-up valence band. Therefore, the strong correlation U likes to pull GdBiTe$_3$ from the topologically non-trivial QAH phase to the topologically trivial ferromagnetic insulator phase. But the energy gap is much more sensitive with the lattice constant than with the correlation U. We have shown  that the SQL GdBiTe$_3$ is close to the quantum critical point between the QAH state and topologically trivial magnetic insulator state, and it is possible to realize the QAH state by tuning the lattice constant with proper substrates. Also the 2D Dirac-type band dispersion shows up at the critical point, shown in Fig.~3c. This is very similar to the case of QSH state at the critical point\cite{bernevig2006d}.

\textbf{ Topological chiral edge states}

The existence of topologically protected chiral edge states is the direct and intrinsic evidence of the QAH phase. It is important to show the explicit features of these topologically protected chiral edge states.  In order to calculate the edge states, we employ the tight-binding method based on maximally localized Wannier functions (MLWF), developed by Vanderbilt and his co-workers\cite{marzari1997}, in the framework of {\it ab-initio} calculations. The edge with Te1$'$ and Bi terminated  along the [11] direction is chosen to show the edge states, shown in the inset of Fig.~4b. Because the topological nature is completely determined by its bulk electronic structure, here we ignore the edge reconstruction of the atoms on the edge, and also make another approximation that the MLWF hopping parameters close to the edge are the same to ones of the bulk. The bulk MLWF hopping parameters are obtained from the {\it ab-initio} calculations of the SQL free-standing GdBiTe$_3$ slab. Similar to the method described in our previous work\cite{zhang2009}, we use an iterative method\cite{sancho1985} to obtain the edge Green's function of the semi-infinite system with an edge along [11] direction. The local density of states (LDOS) can be calculated with the imaginary part of these edge Green's functions, shown in Fig.~4a,b. The LDOS in Fig.~4a shows two edge states for topologically non-trivial GdBiTe$_3$ with the experimental lattice constant $a_0$. One topologically trivial edge state only stays within the valence bands, while the topologically non-trivial edge state ties the conduction band with the valence band. This topologically non-trivial edge state is chiral, and its Fermi velocity in the energy gap is $v_F\simeq3.1\times10^5m/s$ (or $3.86eV\cdot Bohr$). The edge states of topologically trivial ferromagnetic GdBiTe$_3$ with the lattice constant $a=1.02a_0$ are shown in Fig.~4b. Both of these two topologically trivial edge states only stay within the valence bands, and do not connect the valence and conduction bands.

\textbf{ Low-energy effective model}

As the topological nature is determined by the physics near the
$\Gamma$ point for this material, it is possible to write down a
four-band effective Hamiltonian to characterize the low-energy
long-wavelength properties of the system. Starting from the four
low-lying states
$|B_-,-\frac{1}{2}\uparrow\rangle=c_1|{p_x-ip_y},Te1,\uparrow\rangle+c_2|{p_x-ip_y},Bi,\uparrow\rangle$,
$|B_+,\frac{1}{2}\downarrow\rangle=c_1|{p_x+ip_y},Te1,\downarrow\rangle+c_2|{p_x+ip_y},Bi,\downarrow\rangle$,
$|T_+,\frac{3}{2}\uparrow\rangle=d_1|{p_x+ip_y},Te2,\uparrow\rangle+d_2|{p_x+ip_y},Te1',\uparrow\rangle$
and
$|T_-,-\frac{3}{2}\downarrow\rangle=d_1|{p_x-ip_y},Te2,\downarrow\rangle+d_2|{p_x+ip_y},Te1',\uparrow\rangle$
at the $\Gamma$ point, where $c_1$, $c_2$, $d_1$ and $d_2$ are
constants and $\pm\frac{1}{2}(\frac{3}{2})$ are the angular momenta
in $z$ direction. In addition, the ``$\pm$'' subscripts are used to
present the $z$ component of orbital angular momentum $l_z=\pm1$.
Such a Hamiltonian can be constructed by the theory of invariants
for the finite wave vector \textbf{k}. On the basis of the symmetries of the
system, the generic form of the $4 \times 4$ effective Hamiltonian
can be constructed up to the order of $O(k^2)$, and the parameters
of the Hamiltonian can be obtained by fitting to {\it ab initio}
calculations. First we consider the Te-La-Te-Bi-Te
system without magnetic order. The important symmetries of this
system are (1) time-reversal symmetry $T$, (2) reflection symmetry
$\sigma_{x}$ with $x$ axis as the normal axis, and (3) three-fold
rotation symmetry $C_3$ along the $z$ axis. In the basis of
$|B_+,\frac{1}{2}\downarrow\rangle$,
$|T_+,\frac{3}{2}\uparrow\rangle$,
$|B_-,-\frac{1}{2}\uparrow\rangle$ and
$|T_-,-\frac{3}{2}\downarrow\rangle$, the representation of the
symmetry operations is given by $T=\mathcal{K}\cdot
i\sigma^y\otimes\tau^z$, $\sigma_{x}=-i\sigma^x\otimes I$ and
$C_3=exp(-i\frac{2}{3}\pi J_z)$, where $\mathcal{K}$ is the complex
conjugation operator, $\sigma^{x,y,z}$ and $\tau^{x,y,z}$ denote the
Pauli matrices in the spin and orbital space, respectively and $J_z$
is the angular momentum in $z$ direction. For low energy effective
model, we ignore the in-plane anisotropy and impose continuous
rotational symmetry $R(\theta)=exp(-i\theta J_z)$. By requiring
these three symmetries, we obtain the following generic form of the
effective Hamiltonian:
\begin{equation}
\mathcal{H}_0 = \left(
\begin{array}{cccc}
\mathcal{M}(\mathbf{k}) & Ak_+ & -i\Delta_e k_- & A'k_-^2\\
Ak_- & -\mathcal{M}(\mathbf{k}) & A'k_-^2 & 0\\
i\Delta_e k_+ & A'k_+^2  & \mathcal{M}(\mathbf{k}) & Ak_-\\
A'k_+^2 & 0 & Ak_+ & -\mathcal{M}(\mathbf{k})
\end{array}
\right)+\epsilon_0(\mathbf{k})
\end{equation}\label{3D}
with $k_{\pm}=k_x\pm ik_y$, $\epsilon_0(\mathbf{k})=C_0+D
(k_x^2+k_y^2)$, $\mathcal{M}(\mathbf{k})=M_0+B(k_x^2+k_y^2)$. The
block diagonal part is just the BHZ model\cite{bernevig2006d}, and
the off-block diagonal $\pm i\Delta_{e} k_\pm, A'k_{\pm}^2$ terms
breaks the inversion symmetry. Next we add the ferromagnetic
exchange term which breaks time reversal symmetry,
\begin{equation}
\mathcal{H}_{ex} = \left(
\begin{array}{cccc}
-g_BM &  &  &  \\
       & g_TM &  &  \\
       &      & g_BM &  \\
       &  &  & -g_TM
\end{array}
\right)
\end{equation}\label{3D} with $g_B$ and $g_T$ as Land$\acute{e}$-$g$ factors for two different orbitals and $M$ the exchange field introduced by the FM ordering of Gd with half-occupied \textbf{f} electrons. Then the full Hamiltonian can be written as
$\mathcal{H}=\mathcal{H}_{0}+\mathcal{H}_{ex}$. By fitting the
energy spectrum of the effective Hamiltonian with that of the {\it
ab initio} calculations, the parameters in the effective model can
be determined. For GdBiTe$_3$, our fitting leads to $M_0=27meV$,
$A=2.1eV\AA$, $B=7.1eV \AA^2$, $C_0=-14meV$, $D=4.9eV\AA^2$,
$A'\approx 0$, $\Delta_e=1.1eV\AA$, $g_BM=18meV$ and $g_TM=59meV$,
which agree well with the {\it ab initio} results. Such an effective
model can be used for illustration of the formation of the
GdBiTe$_3$ QAH system. As shown in Fig.~2c-e, when we have no FM
ordering, the fact that $M_0$, $B>0$ suggests it is the
topologically trivial insulator as SQL LaBiTe$_3$, and the lack of
inversion symmetry splits the double degeneracy except at $\Gamma$
point. As we adiabatically increase the exchange field $M$, the
$|B_+,\frac{1}{2}\downarrow\rangle$ and
$|T_+,\frac{3}{2}\uparrow\rangle$ states start to move towards each
other. At the transition point, those two bands form a 2D Dirac
cone, and as we further increase the exchange field, those two bands
become inverted and the $Ak_\pm$ term hybridizes them and creates a
band gap again. In this process, a topological phase transition of
SQL GdBiTe$_3$ is shown clearly from a topologically trivial phase
to the QAH phase.

Our theoretical calculations show that a SQL of GdBiTe$_3$ is a 2D stoichiometric magnetic topological insulator, realizing the long-sought-after QAH state. This topological material can be grown by MBE, or by exfoliating the bulk crystal. In addition, YBiTe$_3$ with the same structure is found to be a normal insulator\cite{yan2010}, and could serve as the best substrate for MBE growth. Experimentally, the best way to see the QAH effect is to measure the four-terminal Hall conductance as a function of gate voltage. A quantized plateau in Hall conductance should be observed when the chemical potential is inside the gap. These intrinsic QAH materials could be used for applications with dissipationless electronic transport.

\begin{figure}
   \begin{center}
      \includegraphics[width=4in,angle=-90]{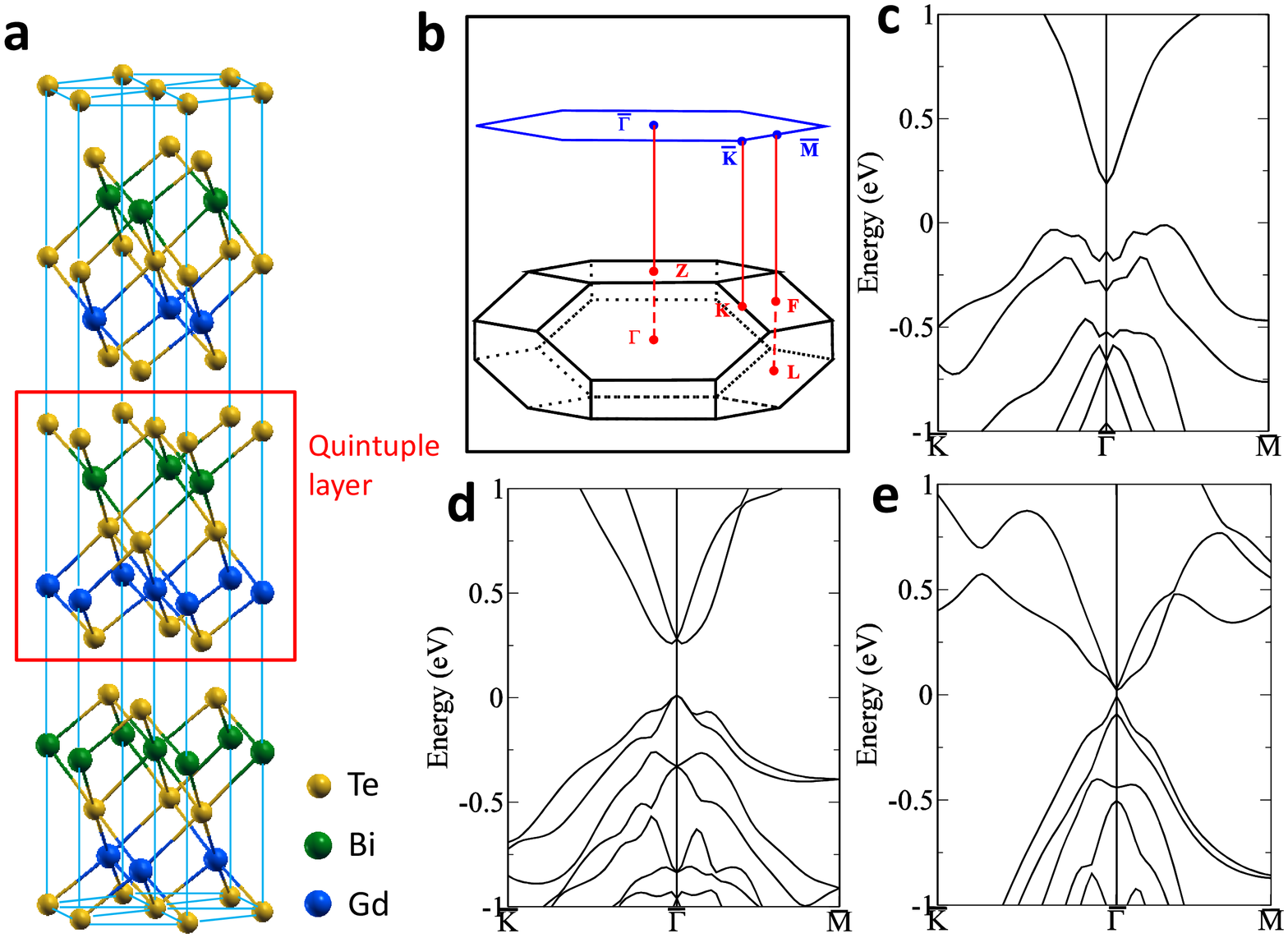}
    \end{center}
    \caption{ {$\mid$\bf Crystal Structure, Brillouin zone and band structure.} {\bf a}, Bulk crystal structure of GdBiTe$_3$. A quintuple
     layer(QL) with $Te1-Bi-Te2-Gd-Te1'$ is indicated by the red box.
     {\bf b}, Brillouin zone for GdBiTe$_3$ with space group $R\overline{3}m$, which has four inequivalent  time-reversal-invariant points $\Gamma(0,0,0)$, $L(\pi,0,0)$,$F(\pi,\pi,0)$ and $Z(\pi,\pi,\pi)$. The projected(111) 2D Brillouin zone is marked by the blue hexagon with its high-symmetry {\bf k} points $\overline{\Gamma}$, $\overline{K}$ and $\overline{M}$. {\bf c, d, e}, Band structure with spin-orbit coupling(SOC) for {\bf (c)} single QL Bi$_2$Te$_3$, {\bf (d)} single QL LaBiTe$_3$ and {\bf (e)} single QL GdBiTe$_3$. The Fermi level is fixed at $0eV$. }
    \label{fig:crystal}
\end{figure}

\begin{figure}
   \begin{center}
      \includegraphics[width=4.5in,angle=-90]{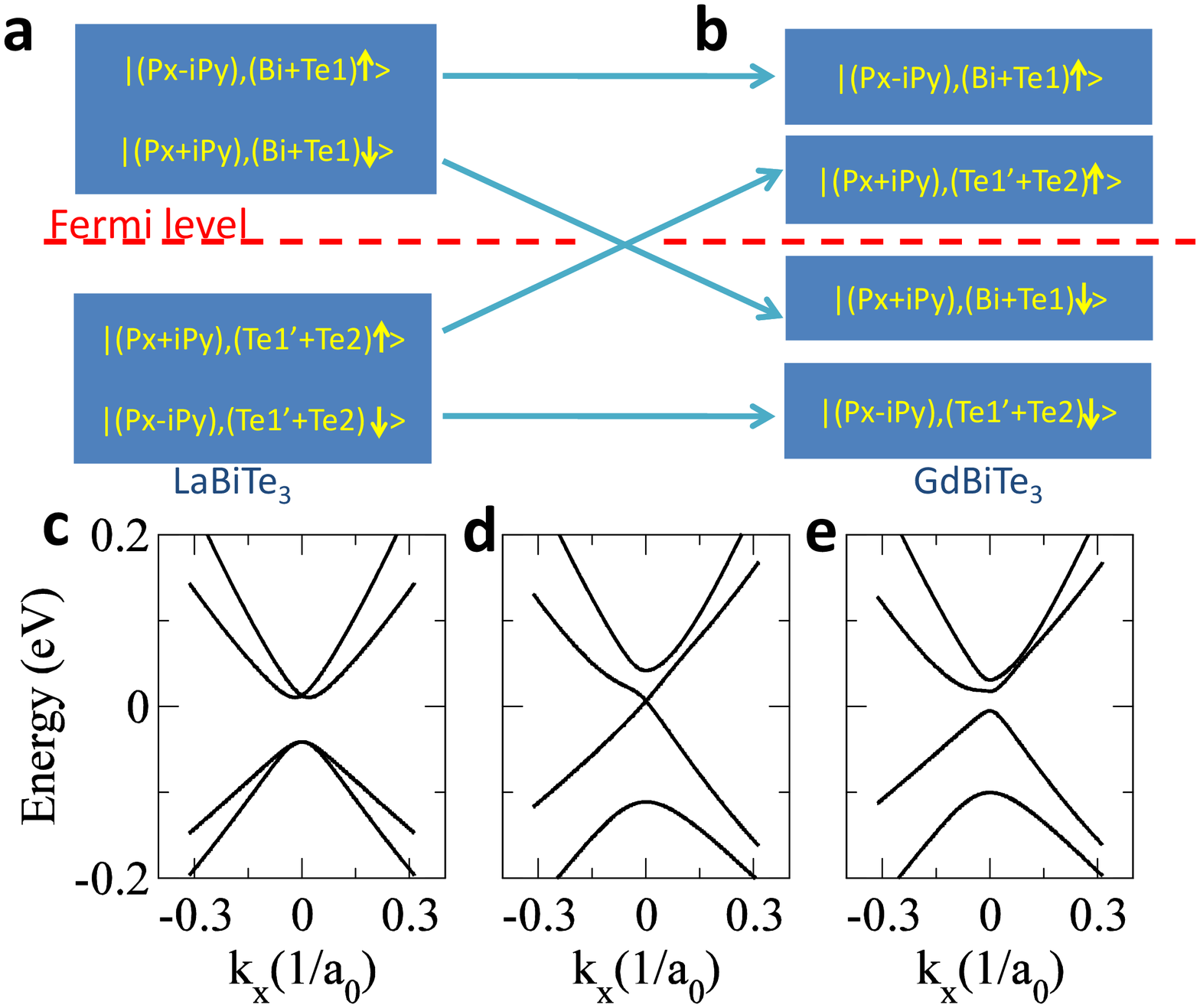}
    \end{center}
    \caption{ {$\mid$\bf Schematic representation of the topological phase transition.} {\bf a,b}, The phase transition from {\bf (a)} the topologically trivial insulator phase of LaBiTe$_3$ to {\bf (b)} the topologically non-trivial QAH phase of GdBiTe$_3$. Because of the Kramers degeneracy,
    $|{p_x-ip_y},{Bi+Te1},\uparrow\rangle$ and $|{p_x+ip_y},{Bi+Te1},\downarrow\rangle$ states at the bottom of the conduction band, as well as states $|{p_x+ip_y},{Te1'+Te2},\uparrow\rangle$ and $|{p_x-ip_y},{Te1'+Te2},\downarrow\rangle$ at the top of the valence band are double degenerated for LaBiTe$_3$. The time-reversal symmetry is broken for GdBiTe$_3$ because of the ferromagnetism of Gd with the half filled \textbf{f} bands. The Kramers degeneracy at $\Gamma$ is removed. Due to the large SOC and the ferromagnetic moment, the band inversion occurs between $|{p_x+ip_y},{Bi+Te1},\downarrow\rangle$ and $|{p_x+ip_y},{Te1'+Te2},\uparrow\rangle$. {\bf c,d,e}, Phase transition based on the Low-energy effective model. The Kramers degeneracy at $\overline{\Gamma}$ for LaBiTe$_3$ is shown in the band structure {\bf (c)} with the time-reversal symmetry, but the Kramers degeneracy for the band structure of {\bf (e)} GdBiTe$_3$ is broken due to the lack of the time-reversal symmetry. The gapless Dirac-type dispersion is shown in {\bf (d)} the band structure at the phase transition point.}
    \label{fig:schematic}
\end{figure}

\begin{figure}
   \begin{center}
      \includegraphics[width=6in,angle=-90]{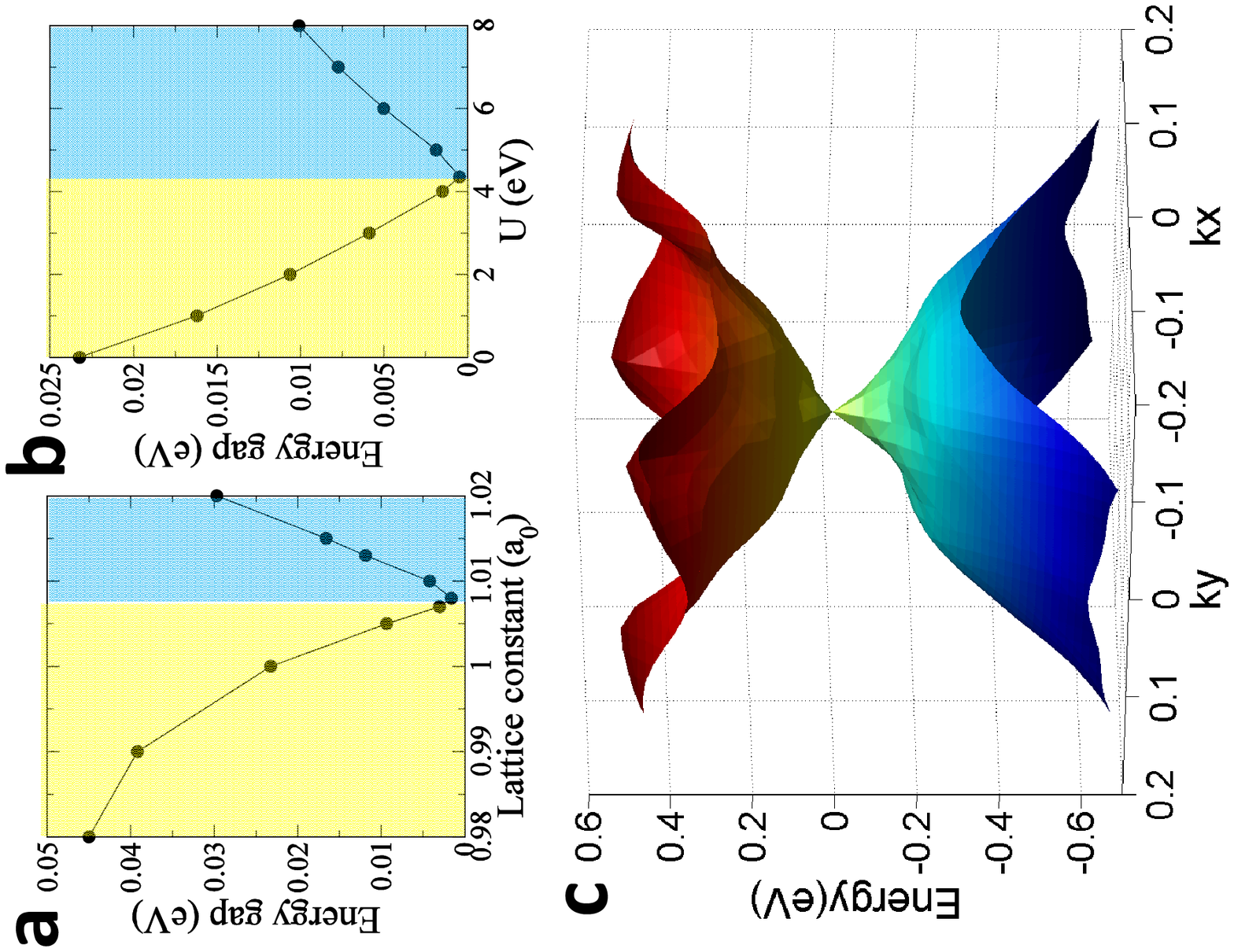}
    \end{center}
    \caption{ {$\mid$\bf Phase diagram.} {\bf a}, The topological phase diagram depending on the lattice constant. $a_0$ is GdBiTe$_3$'s experimental bulk lattice constant. The system is in topologically non-trivial QAH phase with the small lattice constant($a<1.008a_0$) marked by yellow, and it becomes topologically trivial ferromagnetic insulator with the large lattice constant($a>1.008a_0$) marked by blue. {\bf b}, The topological phase diagram depending on the correlation U with fixed experimental lattice constant $a_0$. The GdBiTe$_3$ is in QAH phase with small U($<4.3eV$), and it changes to the topologically trivial ferromagnetic insulator phase with large U($>4.3eV$). {\bf c}, The 2D Dirac-type dispersion of the lowest conduction band and the highest valence band at the transition point.}
    \label{fig:phase}
\end{figure}

\begin{figure}
   \begin{center}
      \includegraphics[width=6in,angle=0]{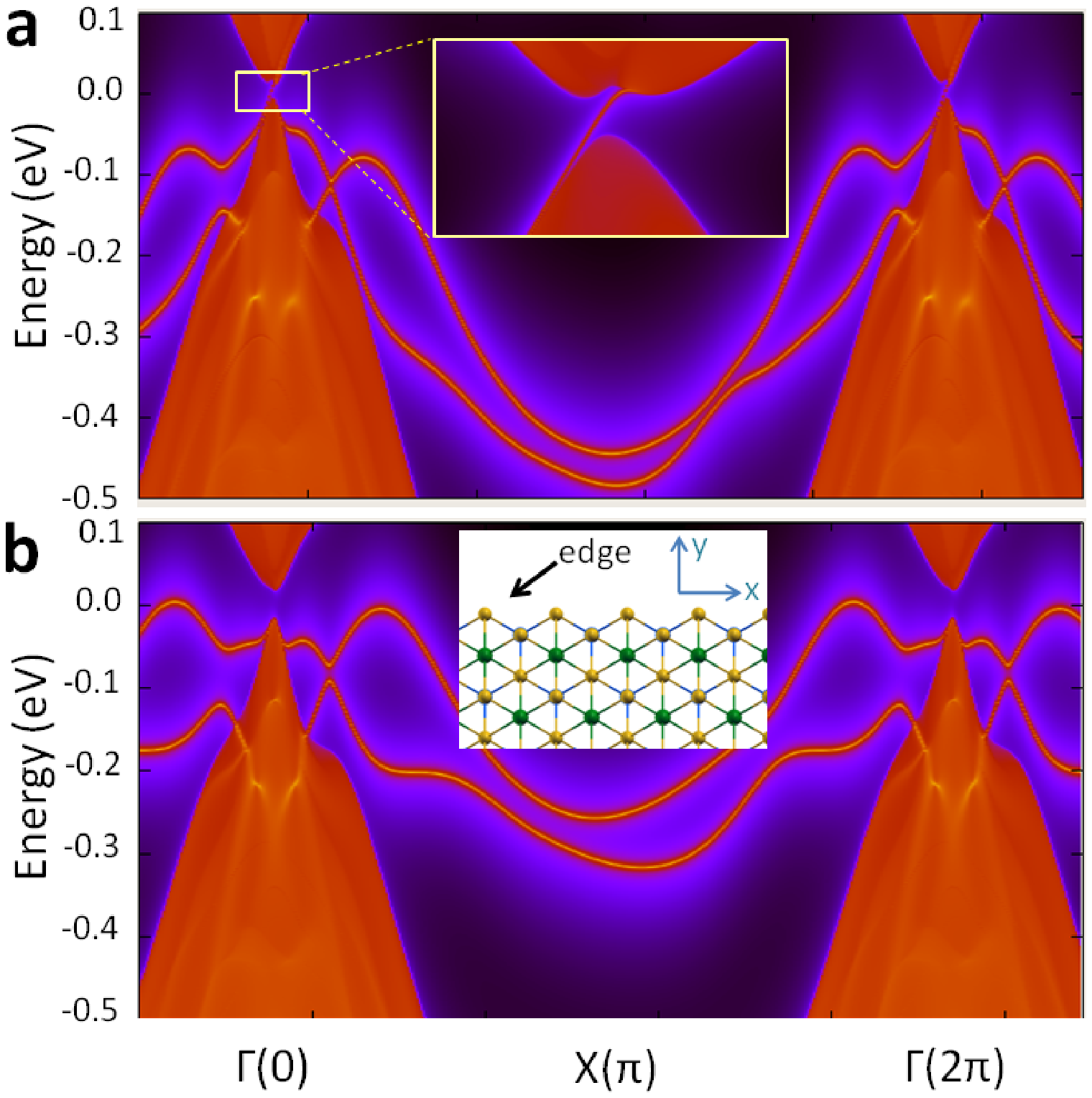}
    \end{center}
    \caption{ {$\mid$\bf Edge states} {\bf a,b}, Energy and momentum dependence of the LDOS for GdBiTe$_3$ with {\bf (a)} QAH phase, and with {\bf (b)} topologically trivial ferromagnetic insulator phase. The red regions indicate the 2D bulk bands and the dark blue regions indicate the bulk energy gap. The edge state are clearly shown in the bulk energy gap. The edge states in {\bf (a)}, which connect the bulk conduction bands and the bulk valence bands, are chiral, and the detailed dispersions around $\overline{\Gamma}$ are zoomed in in the inset. Comparing with the case of QAH, {\bf (b)} the topologically trivial ferromagnetic insulator phase has no chiral edge states. The edge of the single QL is taken to be along $x$ direction, which is the normal axis of the reflection symmetric plane, shown in the inset of {\bf (b)}.}
    \label{fig:edge_state}
\end{figure}

\newpage


%
%
%

\bibliographystyle{Science}


\begin{addendum}
\item We are indebted to B.H. Yan at University of Bremen, X.L. Qi and Q.F. Zhang at Stanford University for their great help. We would like to thank Y.L. Chen and D.S. Kong for their useful discussion. This work is supported by the Army Research Office (No.W911NF-09-1-0508) and the Keck Foundation.

\item[Competing Interests] The authors declare that they have no
competing financial interests.

\item[Correspondence] Correspondence and requests for materials should
be addressed to Shou-Cheng Zhang (email:sczhang@stanford.edu).

\end{addendum}

\end{document}